# AN ONTOLOGY-BASED FRAMEWORK FOR STUDENT PROFILING AND CONTENT PERSONALIZATION IN HIGHER EDUCATION

José Luiz M. Morais (j.morais@unifesp.br)

Prof. Arlindo F. da Conceição (arlindo.conceicao@unifesp.br)

Profa. Cacilda Encarnação Augusto Alvarenga (cacilda.alvarenga@gmail.com )

Profa. Daniela Musa (musa@unifesp.br)

**ABSTRACT**
The expansion of access to Digital Information and Communication Technologies and the offer of distance or semi-distance education courses that make use of virtual learning environments brought changes in the teaching and learning processes, requiring that the student be even more protagonist in this process. The present study aimed to identify important aspects to be considered in the implementation and improvement of self-paced learning and e-learning in higher education courses, with the purpose of rethinking pedagogical models of courses offered at a distance so that they reach even more of your learning objectives. The research is characterized as qualitative, of bibliographic nature, and discusses techniques to monitor and record, electronically and automatically, the results of the process and learning. The importance of processes that store and manage the student's profile is highlighted, both in terms of content and forms of access. The article proposes the use of ontologies to store information about the educational process and presents a computational architecture for this purpose.

## 1 INTRODUCTION

The widespread adoption of Digital Information and Communication Technologies (DICT), together with the expansion of distance education (DE) and blended learning programs that rely on Virtual Learning Environments (VLEs) or e-learning platforms[1] , has transformed teaching and learning processes, instructional methodologies, and teaching practices. These changes have also required students to adopt new roles, becoming increasingly responsible for the self-regulation of their own learning. Self-regulated learning is a psychological process that involves controlling one's thoughts, emotions, and behaviors in ways that promote learning (Zimmerman, 1998). Such behaviors include, for example, effective time management for studying and completing course activities and assignments within established deadlines. DE requires students to exercise an even greater degree of control over their learning process, as they are able to study independently, at the time and place that best suit their individual needs (Pavei & Alliprandini, 2016).

[1] A term used to refer to the teaching and learning process conducted through Virtual Learning Environments (VLEs) or Digital Information and Communication Technologies (DICT), which provide digital instructional content that students can access at the time and place that best suit their needs.

Enrollments in DE undergraduate programs increased substantially between 2008 and 2018. According to the 2018 Higher Education Census, enrollment in DE programs rose by 182.5% during this period, whereas enrollment in face-to-face programs increased by only 25.9%. In 2018, approximately 2 million students were enrolled in DE programs, representing 24.3% of all undergraduate enrollments in Brazil (Brasil, 2019a). It is also important to note that DE has become increasingly integrated into traditional face-to-face programs. For example, Brazilian regulations allow up to 40% of the total instructional workload of an on-campus undergraduate program to be delivered through DE (Brasil, 2019b).

This scenario underscores the need for greater attention to the quality of DE programs, which are increasingly delivered through digital tools and instructional resources, and, consequently, to students' learning outcomes. Although most, if not all, students are familiar with the use of DICT, each student brings a unique personal and academic background, as well as distinct knowledge, competencies, skills, and learning needs.

The standardization of pedagogical models adopted in many DE courses delivered through e-learning or VLEs may compromise the teaching and learning process for students with diverse profiles and learning needs, leading to decreased engagement, course failure, and even dropout. In this context, it is essential to design DE courses that fully leverage the potential of DICT to achieve previously established learning objectives while also embracing different conceptions of learning, such as self-paced learning (SPL), in which students progress through course content at their own pace. The opportunity for students to learn according to their individual pace and needs, and for this flexibility to be incorporated into the planning and development of DE courses, has become increasingly important in the current educational landscape.

According to Jiang et al. (2015), the SPL model is based on the principle that learning curricular content occurs progressively, beginning with simpler or more accessible concepts that gradually evolve into more complex and in-depth knowledge. From this perspective, the authors argue that the curriculum of a DE course, as well as its instructional content, should be organized according to the SPL model in a way that enables students to learn at their own pace, accessing content progressively and building upon previously acquired knowledge that serves both to reinforce existing understanding and to support the construction of new knowledge.

The development of courses that incorporate the SPL approach remains a challenge, as it requires multidisciplinary expertise and practices that may not yet be accessible or feasible for the majority of institutions offering DE courses through e-learning. The present study aims to contribute to the body of knowledge on SPL by presenting techniques for modeling student profiles and personalizing instructional content through an approach that employs ontologies for individual and collective knowledge management.

## 2 IMPORTANT ASPECTS FOR THE IMPLEMENTATION OF SPL AND E-LEARNING

The discussions presented in this section build upon and extend the systematic literature review conducted by Morais and Conceição (2018), which followed the methodology proposed by Kitchenham (2004). The review involved a structured search of the Springer (https://link.springer.com), Science (https://www.sciencedirect.com), IEEE (http://ieeexplore.ieee.org), and ACM (https://dl.acm.org/) databases. The search was conducted using the keywords "self-paced learning," "e-learning," and "higher education" for publications between 2012 and 2018. A total of 89 academic publications of various types (journal and conference papers) addressing the use of SPL in higher education were identified. Of these, 31 studies were considered relevant and were analyzed and categorized according to the countries in which they were conducted, as well as the methodologies and tools they employed (Morais & Conceição, 2018). The review highlighted the main challenges in the field, particularly the difficulties associated with implementing SPL in practice.

This article extends the findings of that review. Considering the main challenges identified, this study proposes and discusses practical approaches to the implementation of SPL and customized e-learning. The proposals presented here are based on the authors' perspective on how new technologies and tools can support this process, particularly by enabling the customization of instructional content.

According to the systematic literature review by Morais and Conceição (2018), the aspects most frequently discussed in studies on DE courses involving e-learning solutions and that may also promote SPL include: (i) flexibility, (ii) collaborative learning, (iii) accessibility, (iv) assessment methods, and (v) feedback strategies.

## 2.1 PROPOSALS FOR ENHANCING THE SPL EXPERIENCE

How can educational support systems, together with teaching and learning practices, address these key aspects? The following sections present proposals aimed at enhancing the SPL learning experience.

2.1 STUDENT PROFILE

In higher education, it is a common practice to maintain students' academic records. However, these records are typically of low granularity, containing little more than the courses completed, grades obtained, and attendance. Although such information may provide general indicators of academic performance, it offers limited insight into what students have actually learned or the knowledge they have acquired.

With the support of appropriate information systems, student profiles can be represented with much greater precision. In addition to completed courses, it is possible to record the exercises students have completed, identify the modules in which they have demonstrated proficiency, and detect gaps in their learning. It is also possible to capture information such as the types of exercises with which students achieve the greatest success, their typical level of dedication to study, their topics of greatest interest, and other variables that can be used to construct a detailed profile of their learning behavior.

Once a comprehensive student profile is available (including knowledge, learning behaviors, and learning goals), it is possible to recommend customized learning pathways that are both more effective and better aligned with each student's objectives.

Understanding a student's profile enables more informed educational decisions and supports the recommendation of appropriate learning pathways. The availability of finer-grained information, with higher level of detail, provides a stronger basis for making pedagogical decisions that are tailored to each learner.

## 2.2 GRANULARITY OF LEARNING OBJECTS

Learning Objects (LOs) are the digital resources used to support the learning process (Wiley et al., 2002). Determining the appropriate level of granularity for the LOs that make up the modules of an e-learning activity remains a challenge. There is no universally accepted standard or guideline, as the optimal level of granularity depends not only on the learning context but also on the characteristics and profiles of the learners.

When course modules are highly granular, that is, when they are composed of smaller LOs, the instructional content is divided into a larger number of discrete topics. This enables the creation of a wider variety of learning pathways tailored to students' specific interests and needs. However, it also increases the risk that important concepts will receive insufficient attention and that the content will become overly fragmented, resulting in learning pathways that lack coherence.

Conversely, when LOs are larger and the level of granularity is lower, learning pathways tend to be more cohesive and better integrated. However, course modules may become lengthy and cognitively demanding, limiting the customization of the teaching and learning process and potentially reducing student engagement.

Thus, identifying an appropriate level of granularity that supports SPL remains an important challenge. How can instructional content be designed to remain both coherent and flexible? The optimal level of granularity may vary according to the knowledge domain, the characteristics of the LOs, the student profile, and the e-learning components used to personalize learning. The prevailing trend is toward the development of increasingly smaller LOs, thereby expanding the potential for customized learning.

## 2.3 MONITORING AND ANALYSIS OF DIGITAL CONTENT USAGE PATTERNS

The use of an e-learning platform makes it possible to monitor patterns of interaction with instructional content. This can be achieved through the analysis of log files, that is, records of users' interactions with the platform and its learning resources. Such monitoring enables the identification of the most frequently accessed content, the pages most often visited, and students' patterns of course access over time, for instance. These insights facilitate the identification of student profiles and, consequently, the refinement of instructional materials to better address their needs. Furthermore, the analysis of LO usage provides valuable evidence regarding the strengths and weaknesses of instructional content.

## 2.4 MEDIA INTEGRATION

Media integration provides a means of diversifying and expanding the instructional content available to students without requiring them to use multiple platforms. Instead, the learning platform can serve as a centralized environment that provides access to and integrates content from different media sources. The underlying idea is to enable students to extend their learning into digital environments they already use regularly, such as social media, while connecting new knowledge to familiar contexts, thereby promoting more meaningful and contextualized learning. Papo (2001) and Tess (2012) likewise emphasize the importance of this type of integration. To be pedagogically effective, however, media integration should always be aligned with the previously established learning objectives.

## 2.5 COLLABORATIVE CONTENT CREATION TOOLS

Collaborative content creation tools are valuable not only for promoting the principles of SPL but also for supporting the long-term sustainability of the learning platform by enabling students themselves to contribute to the creation, revision, and updating of instructional content.

An example of such a collaborative environment is Scratch, a block-based programming platform developed by the Massachusetts Institute of Technology (MIT), primarily intended for children. Within the platform, users can create programming projects and make them publicly available or share them exclusively with the community of registered members. The source code of these projects remains accessible, allowing other users to examine how they were developed, learn from them, and, if desired, reuse or adapt them in their own projects, provided that appropriate credit is given to the original author.

## 2.6 ARTIFICIAL INTELLIGENCE

Artificial Intelligence (AI) is an important technological resource that has the potential to significantly advance the implementation of SPL. AI techniques can be incorporated into learning support systems to make them more intelligent, that is, capable of collecting information about learners, processing that information, and autonomously using or providing it in ways that enhance the learning experience.

In the context of SPL, AI is expected to support students anytime and anywhere, making the learning process less dependent on continuous assistance from tutors. However, AI is not viewed as a replacement for instructors. On the contrary, instructors will continue to play an essential role in designing and planning learning experiences. At the same time, they can use AI to gain meaningful insights into the learning process, such as identifying the most frequent student questions and better understanding how learners acquire new knowledge.

### 2.6.1. Example of an AI Application: A Chatbot to Support Teaching and Learning

From a functional perspective, a chatbot is essentially a conversational interface that enables the exchange of messages between the user and the system. Chatbots are typically anthropomorphized so that users interacting with the system have the impression of communicating with another person. Through this type of interface, it is possible to assess learners' understanding of instructional content in a more natural and fluid manner. As such, chatbots represent a promising interface for future learning systems designed to support the principles of SPL.

## 2.7 ADAPTIVE CONTENT

One of the key requirements for the effective implementation of SPL is content adaptability. When students enroll in a DE course organized according to a traditional, instructor-centered, and standardized instructional model, they are expected to follow a single learning pathway that sequentially covers all topics and modules within predetermined time frames. In other words, the learning trajectory is linear. In contrast, SPL allows each student to progress at their own pace, and one way to enable this customized learning experience is by making instructional content adaptable, that is, by providing multiple learning pathways that students can follow throughout the course.

To make this possible, content recommendation becomes a particularly important feature, as it can suggest learning pathways that are better suited to each student's profile and needs. In VLEs, this functionality can be implemented through mechanisms such as chatbots or electronic questionnaires, allowing students to provide information that serves as the basis for content adaptation. Although content adaptability has been discussed in previous studies, including those by Resende and Dorça (2015), Vialardi et al. (2009), and Primo (2013), this functionality was not identified in any of the 31 studies analyzed in the systematic literature review conducted by Morais and Conceição (2018).

Furthermore, assessment and feedback should be continuous and provided as promptly as possible by both the learner and the system, since these data are essential for evaluating students' progress and the effectiveness of the recommended learning pathway.

## 2.8 FEEDBACK TIMELINESS

To keep students engaged throughout the SPL process, timely feedback from both instructors and VLEs is essential (Lennon, Abbott, & McIntosh, 2015). Continuous feedback was identified as one of the main advantages of using e-learning to support SPL in the systematic literature review by Morais and Conceição (2018). This benefit was highlighted, for example, in the studies by Gitonga, Maina, and Onyango (2016), Lennon, Abbott, and McIntosh (2015), and Schulz, Isabwe, and Reichert (2015). These studies emphasize not only the importance of providing students with prompt feedback on their learning activities and assessments, but also the feasibility of delivering such feedback efficiently and automatically through e-learning technologies.

## 2.9 EFFICIENCY OF FLEXIBILITY

Flexibility was the most frequently cited advantage among e-learning approaches designed to support SPL (Morais & Conceição, 2018), as one of the defining characteristics of this learning model is that it allows students to access instructional content whenever and wherever they choose. However, an important question remains: to what extent is flexibility truly conducive to effective learning?

The very factor that makes flexibility advantageous also poses its greatest challenge: the continuous availability of content. Students must demonstrate self-discipline and commitment so that the freedom to study at any time does not encourage procrastination, thereby reducing learning productivity. To address this challenge, it is important to encourage regular and consistent engagement with course materials. This may be achieved, for example, through frequent updates to instructional content or by rewarding regular access and interaction with learning resources using a gamification approach. Gamification refers to the use of game design elements in non-game contexts (Nacke & Deterding, 2017). In the educational context, this involves engaging students in challenging activities and rewarding their participation or performance. A more restrictive alternative would be to require students to access course content a minimum number of times within a given period; however, such a measure would reduce the benefits associated with flexibility.

The flexibility of both instructional content and access to it enables students with other commitments, such as employment, which is common among higher education students, to organize their schedules and study whenever it is most convenient or feasible. It is therefore the responsibility of instructors and educational institutions to adopt strategies that preserve the effectiveness of flexibility, ensuring that it enhances, rather than hinders, student productivity.

## 2.10 COLLABORATIVE LEARNING

Collaborative learning was the second most frequently cited strength identified in the systematic literature review by Morais and Conceição (2018), as reported in studies such as Herath, Thelijjagoda, and Gunarathne (2015), Ghavifekr and Mahmood (2017), Okaz (2015), Amandu, Muliira, and Fronda (2013), Lennon, Abbott, and McIntosh (2015), Bralić and Divjak (2018), and Gitonga, Muuro, and Onyango (2016). However, the most frequently reported challenge among the analyzed studies was the limited interaction among students in e-learning environments, highlighting the difficulty of fostering the exchange of ideas, information sharing, and the collaborative construction of knowledge.

The incorporation of synchronous e-learning tools, such as online chat systems, can bring students closer together and encourage the real-time exchange of information (Ghavifekr & Mahmood, 2017). Although this approach helps promote a "digital dialogue," it is important to recognize that students are not always available to be online simultaneously. Consequently, the challenge also extends to situations in which synchronous e-learning is not feasible.

One strategy for promoting the collaborative construction of knowledge, even when real-time interaction is not possible, is peer assessment, in which students evaluate the work of their classmates (Chism, 1999). This practice encourages interaction among learners while allowing them to contribute to one another's learning. For peer assessment to be effective, students must recognize the importance of the evaluation process and feel comfortable identifying weaknesses, offering suggestions for improvement, and acknowledging the strengths of their peers' work. To reduce potential bias, each assignment should be evaluated by more than one student (by groups of three or four reviewers, depending on class size, for instance), or the instructor may also participate in the evaluation process, providing a benchmark against which the peer assessments can be compared. Anonymous peer assessment is another strategy for minimizing bias; however, it does not promote student interaction to the same extent.

Peer assessment should not be treated as a one-time activity that ends once a student receives initial feedback from a classmate. Instead, it should be conceived as a continuous and iterative process, allowing students to respond to reviewers' comments and engage in ongoing dialogue about the work. Such reciprocal interaction characterizes a genuine collaborative construction of knowledge that benefits everyone involved in the assessment process.

At a more advanced level of learner autonomy and independence, students themselves can contribute to the creation of instructional content and even entire courses, resulting in the collaborative production of LOs. In this approach, the learning process begins during the design and development of the instructional materials themselves, representing a broader and more comprehensive form of collaboration than the strategies discussed previously. For collaborative content production to be successful, students must demonstrate commitment and responsibility, as well as possess sufficient prior knowledge of the subject matter, since they become co-responsible for designing the learning pathways offered by the course. Consequently, the greater the diversity of expertise and academic backgrounds among students, the richer and more comprehensive the instructional content becomes (Page, 2010).

## 2.11 REDEFINING THE ROLE OF THE INSTRUCTOR

Despite the emphasis on learner autonomy in SPL, instructors continue to play a fundamental role in the learning process. The lack of instructor feedback has been identified as one of the major challenges reported in studies such as those by Maina and Kihoro (2017) and Asoodar, Vaezi, and Izanloo (2016), cited in the systematic literature review by Morais and Conceição (2018), and must be addressed to support student success.

The term *self-paced learning* is often mistakenly associated with the absence of the instructor; however, this interpretation does not reflect the nature of the approach. Instructors should remain available to answer students' questions, provide timely feedback, and support them in making informed learning decisions. Although each learner progresses at an individual pace, it is the instructor's responsibility to help students discover that pace, develop the judgment needed to choose appropriate learning pathways, and recommend relevant instructional resources. This role extends throughout the entire educational process: before the course, through the design and organization of instructional materials; during the course, by recommending additional learning resources; and after the course, by supporting the continuation and deepening of students' learning.

Ideally, instructors should maintain their educational role not only throughout the course but also after its completion. In the context of higher education, for example, instructors may continue engaging with former students by keeping them on mailing lists through relevant materials and discussions are, or by creating groups on social media platforms to facilitate ongoing communication, knowledge exchange, and content updates.

## 2.12 STUDENT ENGAGEMENT

Student engagement and commitment are essential for the successful implementation of SPL. According to Hill (2013), participants in free online courses such as Massive Open Online Courses (MOOCs) can be classified into five groups: (i) No-shows, who register for a course but never access it. These learners are estimated to constitute the majority of MOOC registrants; (ii) Observers, who watch videos, read discussions, and follow course activities but do not participate in any form of assessment, demonstrating little commitment to course completion; (iii) Drop-ins, who engage only with specific topics that are relevant to their interests without intending to complete the entire course; (iv) Passive participants, who access the available instructional content but do not actively participate in course activities; and (v) Active participants, who engage fully in the course with the intention of completing it by actively participating in discussions and assessments.

For learning in a DE course, such as a MOOC, to be effective, students must recognize that their active participation is fundamental. Learners should go beyond merely accessing instructional content by committing to complete assessments, contributing to their peers' learning, and providing constructive feedback on both the course and the platform or VLE through which it is delivered.

To encourage a greater proportion of students to become active participants, it is essential that course designers and instructors clearly communicate the importance of learner participation and adopt strategies that foster engagement, such as gamification (Alenezi & Shahi, 2015).

From this perspective, students classified as *Observers*, *Drop-ins*, and *Passive participants* have the potential to become active learners, since they already participate in the course to some extent and primarily need to increase their level of engagement. An example of a gamification element that can encourage sustained participation is the *streak* mechanism implemented in the Duolingo language-learning platform. Each consecutive day that users complete a predetermined number of learning activities increases their streak by one day. Streak data can then be used to create leaderboards and reward systems that recognize and encourage consistent engagement among learners.

## 2.13 GAMIFICATION

Gamification, which involves applying game design strategies to DE courses, is one of the main approaches for maintaining student interest and increasing interaction among learners (Lennon, Abbott, & McIntosh, 2015). However, its implementation must be carefully planned to ensure that the competition it introduces does not become a source of discouragement or demotivation for students (Morais & Conceição, 2018).

One way to incorporate game elements into education is through reward systems integrated into learning platforms and support tools. An example is the badge system adopted by Khan Academy, which rewards students with different types of badges, classified according to their level of difficulty and associated point values, based on the activities they complete. Khan Academy also organizes competitions and challenges among its users, offering awards to the highest-scoring schools and holding in-person ceremonies to recognize the top performers of the academic year.

To prevent competitiveness from negatively affecting student motivation, reward systems should recognize not only individual achievement but also collaborative behavior. For example, leaderboards should not focus exclusively on students' individual performance but should also acknowledge those who are most supportive of their peers, participate most actively, suggest the greatest number of improvements, and contribute the most with new instructional content to the platform.

Thus, gamification serves as a strategy for sustaining student interest and motivation. However, it is important that students remain motivated primarily by the benefits of learning and knowledge construction, rather than by the rewards themselves.

## 2.14 EVALUATING DIGITAL CONTENT THROUGH STUDENT PROFILES

The evaluation of digital instructional content can be a key factor in the success of DE courses based on the principles of SPL. Because the same content is made available to many students, each of whom may use it differently, it is important to evaluate not only the quality of the instructional materials but also how they are used and for what purpose (Chang, Hung, & Lin, 2015). For example, one student may use a proposed activity as practice for an upcoming assessment, whereas another may use the same activity to learn a new concept through the practical exploration of different approaches to solving a problem.

Effective evaluation of instructional content first requires students' commitment, since they are the primary users of these materials and are therefore best positioned to provide feedback regarding their quality and usefulness within the broader context of the course. This feedback can be collected through online questionnaires completed at the end of each module or stage of study.

Figure 1.a illustrates the traditional approach to content evaluation, in which the process is organized as a cycle consisting of instructional planning, the development of learning objects, the implementation of instructional content, and the evaluation of learning outcomes, with the evaluation results serving as input for the next planning cycle. However, this evaluation process can be enhanced by incorporating an additional element that contributes to the customization of learning: student profiles.

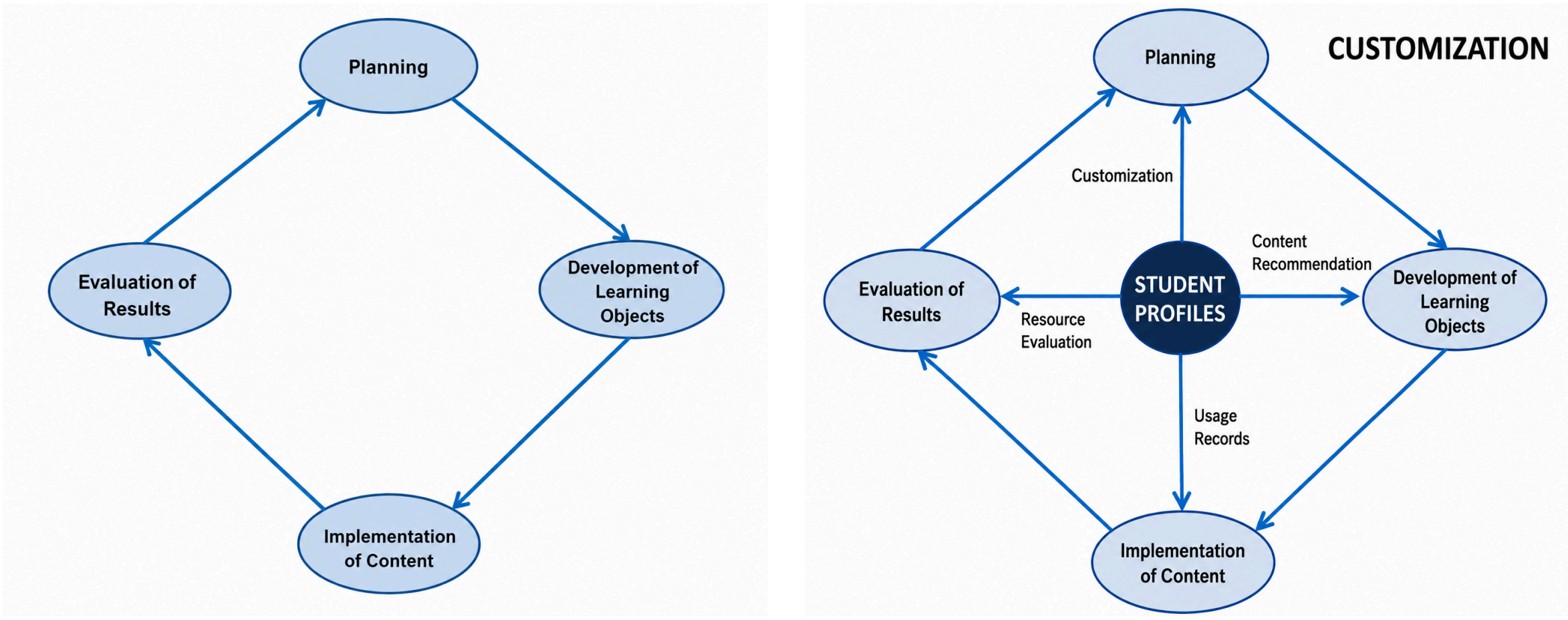

**a.** Traditional content evaluation cycle **b.** Customized cycle

**Figure 1:** Content evaluation cycle

Figure 1.b illustrates a content evaluation process that incorporates student profiles to support content customization. Based on these profiles, it becomes possible to analyze how instructional content has been accessed, recommend new content to students with greater precision and effectiveness, and examine how learners evaluate the available resources and interact with the instructional materials provided. The work of Sein-Echaluce, Fidalgo-Blanco, and García-Peñalvo (2017) further reinforces the importance of incorporating student profiles into this process.

Therefore, the evaluation of digital instructional content is important not only for improving the quality of the materials made available to students but also for directly supporting the implementation of the SPL approach.

## 3 ONTOLOGIES FOR KNOWLEDGE MANAGEMENT AND CONTENT CUSTOMIZATION

Based on the discussions and proposals presented in the previous section, it is possible to envision the implementation of learning platforms and VLEs that incorporate functionalities designed to address the identified challenges and support the implementation of SPL.

As an extension of this work, we propose a student profile management architecture based on ontologies. In this context, ontologies are used as a means of representing knowledge and formally specifying the relationships among concepts through classes, individuals, and properties (Noy & McGuinness, 2001). Their purpose is to organize knowledge in a structured manner, facilitating the identification of the knowledge already acquired by students, as well as their knowledge gaps, while enabling the future recommendation of LOs to create customized learning pathways.

The proposed architecture comprises four main components: the browser, which serves as the user interface; web services; the repository; and the ontologies. The repository, which can be implemented using Apache Spark, is the architectural component responsible for storing knowledge-related information, including both the student's knowledge and the knowledge associated with a given domain. It stores the ontology individuals representing knowledge, which are compared to determine what the student has already mastered and to identify existing knowledge gaps.

For ontology implementation, we propose the use of the Web Ontology Language (OWL), a language developed for applications that require not only knowledge representation but also knowledge processing. OWL provides an extended vocabulary for describing classes, properties, and relationships, while supporting formal semantics (McGuinness & van Harmelen, 2004).

To query and manipulate the ontologies, we propose the use of SPARQL, a query language designed to retrieve and manipulate information stored in Resource Description Framework (RDF) graphs, while also providing support for OWL ontologies. The Apache Jena API can be employed to programmatically manipulate the ontologies using SPARQL syntax. Because Apache Jena supports OWL and SPARQL, it also enables integration with the Apache Spark-based repository.

## 3.1 THE PROPOSED ONTOLOGY

To represent knowledge, we developed an ontology named ***CourseStructure***, which models the structure of an educational program, whether a higher education program or another type of academic program, covering elements ranging from academic disciplines to the learning objects used within each instructional module.

The ontology consists of six classes:

(i) ***Institution***, which represents the institution offering the educational program. An *Institution* is associated with individuals of the *Program* class, representing the programs it offers;

(ii) ***Program***, which represents a structured area of knowledge. A *Program* corresponds to an academic program offered by an institution and comprises multiple disciplines;

(iii) ***Discipline***, which represents an academic discipline belonging to a *Program*. A *Discipline* contains instructional modules and is also related to other individuals of the same class to represent prerequisite disciplines;

(iv) ***Module***, which represents the instructional modules that compose a *Discipline*. Each *Module* contains learning objects and, similarly to the *Discipline* class, may be associated with other *Module* individuals to represent prerequisite modules;

(v) ***Test***, which represents the assessment instruments associated with the modules of a discipline; and

(vi) ***LearningObject***, which represents the smallest instructional unit in the ontology, namely the learning objects themselves. A *LearningObject* may be shared across multiple modules, disciplines, or academic programs.

Figure 2 illustrates the proposed ontology.

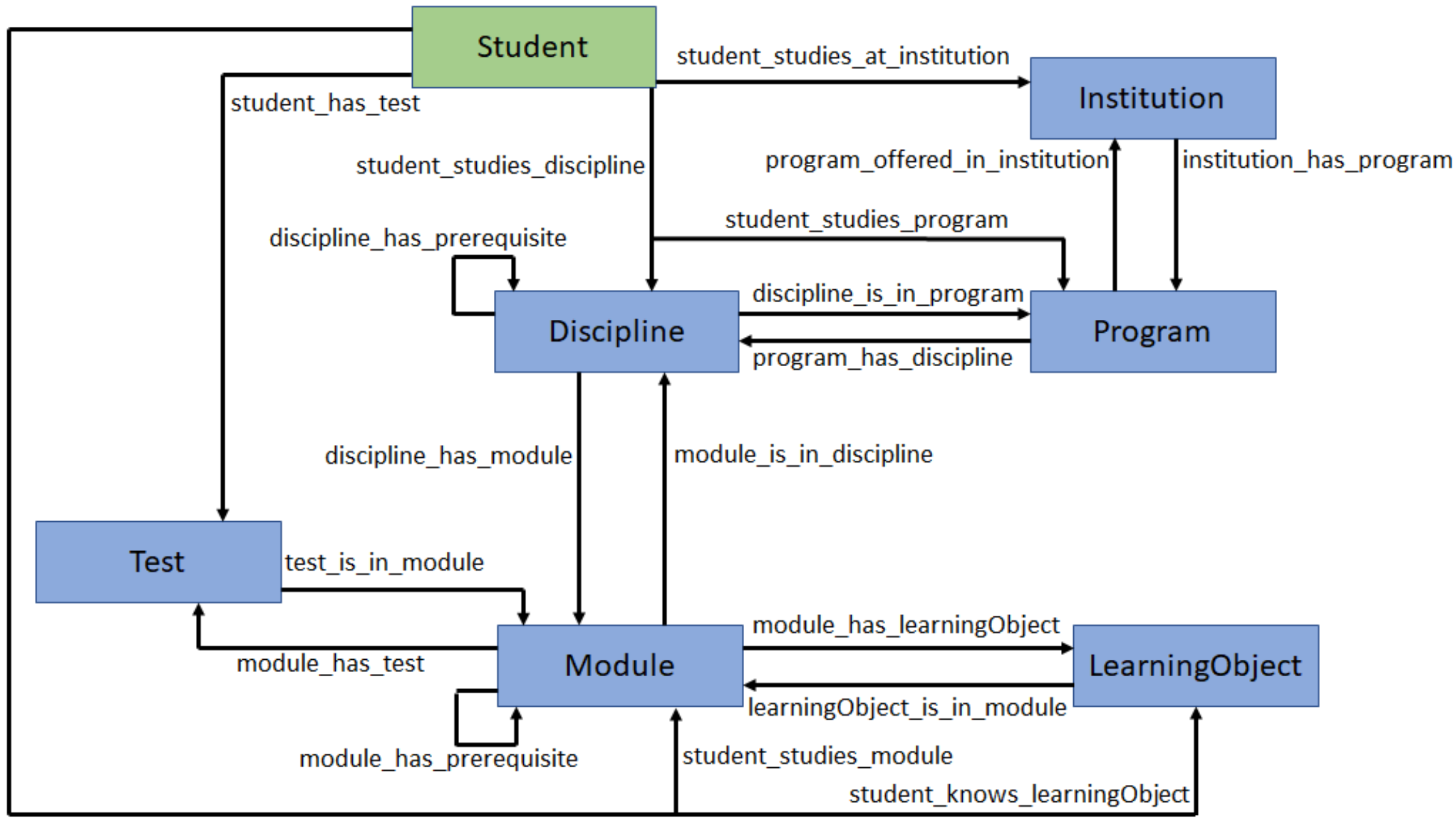

**Figure 2:** Proposed ontology for knowledge representation.

This ontology makes it possible to map all the knowledge represented within an educational program and, based on that representation, generate customized learning pathways according to each student's profile, as illustrated in Figure 1.b. In this model, each student's profile includes a checklist that is progressively updated as the student advances through the course, reflecting the knowledge acquired according to the ontology-based knowledge representation.

## 4 STUDENT PROFILE USE CASES

Based on the systems, methods, and tools discussed throughout this paper, it is possible to envision a scenario that illustrates a future perspective on student learning in the context of higher education, particularly considering the principles of SPL.

Upon completing secondary education, a student's academic record could be analyzed, subject by subject, by a computerized system that combines this information with data regarding the student's primary academic interests, collected, for example, through online questionnaires. Based on this analysis, the system could generate a list of recommended academic programs that best match the student's demonstrated interests, abilities, and academic background.

After reviewing the recommendations and making a choice, the student enrolls in an undergraduate program and receives the curriculum, which includes the mandatory courses required for degree completion in the chosen field. While the curriculum and course syllabi are the same for all students, the learning activities are adapted according to each student's performance and aptitudes, as identified through continuous feedback exchanged between the learner and the VLE via chatbots, user logs, online forms, and assessment questions. Student assessment is continuous and extends beyond traditional examinations. It also takes into account, for example, the instructional resources used by the student, the time required to master each module, and the extent of the student's collaborative participation in the course. At the same time, the learning tools and the academic program itself are continuously evaluated by students, who provide constructive feedback and propose improvements regarding both the instructional content and its presentation.

Throughout the program, students have access to instructional content in a variety of formats, including text, videos, scholarly articles, interactive exercises, challenges, projects, and learning activities, all integrated within the VLE. During these activities, students can share their work with their peers, who may evaluate it and provide suggestions for improvement. Students may also collaborate on group projects based on shared interests. In addition to enabling students to share their work, the VLE allows them to contribute new and relevant instructional content that complements the existing learning materials. Upon completing an activity, challenge, or module, or even making a contribution recognized as valuable by the learning community, students receive a reward that is recorded in their profile, contributing to their academic reputation within the educational community.

As students progress through their courses, the platform recommends customized learning pathways based on their individual profiles. These profiles are continuously updated using the information exchanged between the student and the system, together with analyses of the student's performance in previously completed courses. Consequently, students can select the learning pathway that best suits their interests and follow it at their own pace. Because these pathways are customized according to the learner's accumulated history and experiences recorded by the platform, they are expected to provide a more effective learning experience.

Such a system therefore accompanies students throughout their academic journey. From the perspective of computer systems, it should provide four essential characteristics: usability, high availability, interoperability, and security. Usability refers to the ease with which students can interact with the system. Ideally, the platform should automatically update student profiles as learning activities are completed. High availability means that the system should be continuously accessible, as is expected of modern cloud-based platforms. Interoperability is essential to allow a student's profile to be updated by multiple educational institutions throughout their academic career.

Finally, security issues, particularly those related to privacy and identity protection, are of fundamental importance. Information regarding a student's educational background should be treated as sensitive data and protected against unauthorized access. Furthermore, it is possible to envision scenarios in which anonymized and aggregated data could be made available for statistical analyses and educational research while preserving students' privacy.

As future work, we intend to develop computational tools to maintain and update the data models underlying student profiles. Given the computational requirements discussed above, particularly those related to data privacy, one promising direction is to explore the adaptation of the concepts of self-sovereign identity (SSI) and verifiable credentials (Wang & De Filippi, 2019), together with the use of blockchain technology (Bore et al., 2017) as a mechanism for managing student identities and storing student profiles.

## 5 CONCLUSION

Based on the analysis of the challenges inherent in the effective implementation of SPL, together with the proposed techniques for modeling and managing student profiles and personalizing instructional content, this study is expected to contribute to the planning and implementation of courses that better address students' needs. Future work will focus on managing acquired knowledge in a structured manner, supported by computational tools specifically designed for this purpose. Improving the management of the learning process is expected to remain one of the major challenges to be addressed in the coming years.

## REFERENCES

ALENEZI, Ahmed Maajoon; SHAHI, Krishna Kirti. Interactive e-learning through second life with blackboard technology. **Procedia-Social and behavioral sciences**, v. 176, p. 891-897, 2015.

AMANDU, Gerald Matua; MULIIRA, Joshua Kanaabi; FRONDA, Dennis Cayaban. Using moodle e-learning platform to foster student self-directed learning: Experiences with utilization of the software in undergraduate nursing courses in a Middle Eastern university. **Procedia-Social and Behavioral Sciences**, v. 93, p. 677-683, 2013.

ASOODAR, Maryam; VAEZI, Shahin; IZANLOO, Balal. Framework to improve e-learner satisfaction and further strengthen e-learning implementation. **Computers in Human Behavior**, v. 63, p. 704-716, 2016.

BORE N; KARUMBA S; MUTAHI J; DARNELL SS; WAYUA C; WELDEMARIAM K. Towards blockchain-enabled school information hub. In Proceedings of the **Ninth International Conference on Information and Communication Technologies and Development**. Nov 16 (pp. 1-4), 2017.

BRALIĆ, Antonia; DIVJAK, Blaženka. Integrating MOOCs in traditionally taught courses: achieving learning outcomes with blended learning. **International Journal of Educational Technology in Higher Education**, v. 15, n. 1, p. 2, 2018.

BRASIL. Instituto Nacional de Estudos e Pesquisas Educacionais Anísio Teixeira (Inep). **Censo da Educação Superior 2018 :** notas estatísticas. Brasília, 2019a.

BRASIL (b). **Portaria Nº 2.117, de 06 de dezembro de 2019.** Disponível em: http://www.in.gov.br/en/web/dou/-/portaria-n-2.117-de-6-de-dezembro-de-2019-232670913 Acesso em: 10 jan. 2019.

CHANG, Ray I.; HUNG, Yu Hsin; LIN, Chun Fu. Survey of learning experiences and influence of learning style preferences on user intentions regarding MOOC s. **British Journal of Educational Technology**, v. 46, n. 3, p. 528-541, 2015.

CHISM, Nancy Van Note. **Peer Review of Teaching. A Sourcebook**. Anker Publishing Company, Inc., 176 Ballville Road, PO Box 249, Bolton, MA 01740-0249, 1999.

DESPUJOL, Ignacio M. et al. Analysis of demographics and results of student's opinion survey of a large scale mooc deployment for the spanish speaking community. In: **2014 IEEE Frontiers in Education Conference (FIE) Proceedings**. IEEE, 2014. p. 1-8.

DORÇA, Fabiano A. et al. An automatic and dynamic approach for personalized recommendation of learning objects considering students learning styles: an experimental analysis. **Informatics in education**, v. 15, n. 1, p. 45, 2016.

GHAVIFEKR, Simin; MAHMOOD, Hazline. Factors affecting use of e-learning platform (SPeCTRUM) among University students in Malaysia. **Education and Information Technologies**, v. 22, n. 1, p. 75-100, 2017.

GITONGA, Rhoda; MUURO, Maina; ONYANGO, George. Technology integration in the classroom: A case of students experiences in using Edmodo to support learning in a blended classroom in a Kenyan University. In: **2016 IST-Africa Week Conference**. IEEE, 2016. p. 1-8.

HERATH, Chaminda P.; THELIJJAGODA, Samantha; GUNARATHNE, W. K. T. M. Stakeholders' psychological factors affecting E-learning readiness in higher education community in Sri Lanka. In: **2015 8th International Conference on Ubi-Media Computing (UMEDIA)**. IEEE, 2015. p. 168-173.

HILL, P. Emerging student patterns in MOOCs: A (revised) graphical view. e-Literate. **URL: http://mfeldstein. com/emerging_student_patterns_in_moocs_graphical_view**, 2013.

Jiang, L.; ,Meng, D.; Zhao, Q.; Shan, S. ; Hauptmann, A.G. Self-Paced Curriculum Learning. **Proceedings of the Twenty-Ninth AAAI Conference on Artificial Intelligence**. January 25–30, 2015, Austin Texas, USA.

KITCHENHAM, Barbara. Procedures for performing systematic reviews. **Keele, UK, Keele University**, v. 33, n. 2004, p. 1-26, 2004.

LENNON, Alison; ABBOTT, Malcolm; MCINTOSH, Keith. Chasing higher solar cell efficiencies: Engaging students in learning how solar cells are manufactured. In: **2015 IEEE International Conference on Teaching, Assessment, and Learning for Engineering (TALE)**. IEEE, 2015. p. 267-271.

MAINA, Elizaphan Muuro; KIHORO, John M. Learner experience of e-learning mode in institutions of higher learning: A case of Kenyan Universities. In: **2017 IST-Africa Week Conference (IST-Africa)**. IEEE, 2017. p. 1-9.

MCGUINNESS, Deborah L. et al. OWL web ontology language overview. **W3C recommendation**, v. 10, n. 10, p. 2004, 2004.

MORAIS, José Luiz Machado; DA CONCEIÇÃO, Arlindo Flavio. Ferramentas Tecnológicas e Metodologias de Apoio à Aprendizagem Personalizada no Ensino Superior: uma Revisão Sistemática Technological Tools and Supporting Methodologies for Personalized Learning in Higher Education. **Revista Informática na educação: teoria e prática**, v. 21, n. 3, 2018.

NACKE, Lennart E.; DETERDING, Christoph Sebastian. The maturing of gamification research. **Computers in Human Behaviour**, p. 450-454, 2017.

NOY, Natalya F. et al. Ontology development 101: A guide to creating your first ontology. 2001.

OKAZ, Abeer Ali. Integrating blended learning in higher education. **Procedia-Social and Behavioral Sciences**, v. 186, p. 600-603, 2015.

PAGE, Scott E. **Diversity and complexity**. Princeton University Press, 2010.

PAPO, William. Integration of educational media in higher education large classes. **Educational Media International**, v. 38, n. 2-3, p. 95-99, 2001.

PAVESI, M.A.; ALLIPRANDINI, P.M.Z. Análise da Aprendizagem Autorregulada de Estudantes de Cursos a Distância em Função das Áreas de Conhecimento. **Educação, Formação & Tecnologias,** V. 9; N.1.,2016 p. 3-15.

PRIMO, Tiago Thompsen. Método de representação de conhecimento baseado em Ontologias para apoiar Sistemas de Recomendação Educacionais. 2013.

RESENDE, Daniel Teixeira; DORÇA, Fabiano Azevedo. Recomendação de conteúdo personalizada com base em estilos de aprendizagem: uma abordagem prática. **Revista Brasileira de Informática na Educação**, v. 23, n. 3, p. 11-25, 2015.

SCHULZ, Renée; ISABWE, Ghislain Maurice; REICHERT, Frank. Investigating teachers motivation to use ICT tools in higher education. In: **2015 Internet Technologies and Applications (ITA)**. IEEE, 2015. p. 62-67.

SEAMAN, Julia E.; ALLEN, I. Elaine; SEAMAN, Jeff. Grade Increase: Tracking Distance Education in the United States. **Babson Survey Research Group**, 2018.

SEIN-ECHALUCE, María Luisa; FIDALGO-BLANCO, Ángel; GARCÍA-PEÑALVO, Francisco J. Adaptive and cooperative model of knowledge management in MOOCs. In: **International Conference on Learning and Collaboration Technologies**. Springer, Cham, 2017. p. 273-284.

STEFANSSON, Gunnar; LENTIN, Jamie. From Smileys to Smileycoins: Using a Cryptocurrency in Education. **Ledger**, v. 2, p. 38-54, 2017.

TESS, Paul A. The role of social media in higher education classes (real and virtual)–A literature review. **Computers in human behavior**, v. 29, n. 5, p. A60-A68, 2013.

VIALARDI, César et al. Recommendation in Higher Education Using Data Mining Techniques. **International Working Group on Educational Data Mining**, 2009.

WANG F; DE FILIPPI P. Self-sovereign identity in a globalized world: Credentials-based identity systems as a driver for economic inclusion. **Frontiers in Blockchain**, Jan 23; 2:28, 2020.

WILEY, David A. et al. **The instructional use of learning objects**. Bloomington, IN: Agency for instructional technology, 2002.

ZHANG, Dongsong et al. Can e-learning replace classroom learning?. **Communications of the ACM**, v. 47, n. 5, p. 75-79, 2004.

ZIMMERMAN, B.J. Academic Studying and the Development of Personal Skill: a self-regulatory perspective. **Educational Psychologist**, 1998, 33(2), p. 73-86.